\documentclass[12pt]{article}
\usepackage{amsmath}

\newcommand{\mql}{\frac{\lambda_1}{2m^{}_Q}}
\newcommand{\mqll}{\frac{\lambda_2}{2m^{}_Q}}
\newcommand{\mqlt}{\frac{\lambda_1^{0^+}}{2m^{}_Q}}
\newcommand{\mqls}{\frac{\lambda_1^{1^+}}{2m^{}_Q}}
\newcommand{\mqlls}{\frac{\lambda_2^{1^+}}{2m^{}_Q}}

\newcommand{\cc}{\cos^2\theta_Q}
\newcommand{\sss}{\sin^2\theta_Q}

\begin{document}

\title{\textsf{Mixing Angle of $\varXi_Q-\varXi_Q'$ \\ in Heavy Quark Effective Therory}}
\author{\textsf{Yoshimitsu Matsui }\thanks{matsui@vega.aichi-u.ac.jp} \\
	Faculty of Law, Aichi University,\\Nagoya, Aichi 453-8777, Japan}
	\date{}
\maketitle

\begin{abstract}
The Heavy Quark Effective Theory provides a systematic method 
to estimate a mixing angle of hadron states in 
a heavy quark, such as the charm quark (c) and the bottom quark (b).
By using this method, the mixing angle of the baryons 
$\varXi_Q-\varXi_Q'$ can be estimated.
It is found that the mixing angle between 
$\varXi_Q$ and $\varXi_Q'$ is given by 
 $\theta_b = 4.51^{\circ} \pm 0.79 ^{\circ}$ for $Q = b$ case and
 $\theta_c = 8.12^{\circ} \pm 0.80 ^{\circ}$ for $Q = c$ case.
\end{abstract}

\section{Introduction}

In recent year, the experiment of the baryon containing a heavy quark 
have progressed\cite{LHCb}. 
Especially, many states of the baryon containing a bottom quark 
have been found and those characteristics are clarified \cite{PDG2020}. 
The environment which tests the validity of the Heavy Quark Effective 
Theory (HQET) \cite{Isgur:1989vq,Neubert:1993mb} 
to the baryons containing a bottom quark was ready.

In the 1990s, it was pointed out that theoretical values drawn from the HQET about 
the baryons containing a bottom quark, such as the mass relations \cite{DELPH95}
and the lifetime ratio\cite{Bigi:1992su}-\cite{Ito:1997qq}, 
were discrepant to the experimental values. 
The mass relation of the baryons $\varSigma_b$, $\varSigma^*_b$, $\varSigma_c$, $\varSigma^*_c$, the mesons $B$, $B^*$, $D$, $D^*$, 
charm quark ($c$) and bottom quark ($b$),
\begin{equation}
	\frac{M \left(\varSigma_b^* \right) -M \left(\varSigma_b \right)}{M \left(\varSigma_c^* \right) -M \left(\varSigma_c \right)}
	=\frac{M \left(B^* \right) - M \left(B \right) }{M \left(D^* \right) - M \left(D \right) }
	=\frac{m_c}{m_b}
\end{equation}
was derived by using HQET with $O\left(\Lambda_{QCD} / m_Q \right)$ corrections. 
If the experimental values were 
substituted for the each side, we can estimated as follow.
\begin{eqnarray}
\frac{M \left(\varSigma_b^* \right) -M \left(\varSigma_b \right)}
{M \left(\varSigma_c^* \right) - M \left(\varSigma_c \right)} &=& 0.2994 \pm 0.0034 \\
\frac{M \left(B^* \right) - M \left(B \right) }{M \left(D^* \right) - M \left(D \right) } & = & 0.31996 \pm 0.00065 \\
\frac{m_c}{m_b} & = & 0.3038 \pm 0.0053 
\end{eqnarray}
Since the magnitude of next order corrections can be estimated up to 0.04, 
this mass relation was consistent to the experimet.
The lifetime ratio of the baryon $\varLambda_b$ and the meson $B_d$ can be drived from HQET.
We can estimate the ratio as,
\begin{equation}
	\frac{\tau(\varLambda_b)}{\tau(B_d)} = 0.98 + {\cal O}\left( \Lambda_{QCD}^3/m_b^3 \right) .
\end{equation}
Since the experimental values of this ratio was,
\begin{equation}
	\frac{\tau(\varLambda_b)}{\tau(B_d)} = 0.9684 \pm 0.0065,
\end{equation}
the result of HQET was consistent to the experiment.
It was profoundly interesting that the lifetime ratio was nearer than 
the ratio of hadron mass, $M\left(B_d\right)/M\left(\varSigma_b^* \right)=0.93950 \pm 0.00003$, by one.

The mixing angle for each case of $Q=c$ and $Q=b$, had been estimated by some methods, 
the QCD sum rules\cite{Aliev:2010ra}, the bag model calculations\cite{Bernotas:2008fv} and 
the quark model\cite{Franklin:1981rc,Franklin:1996ve}.
And the mixing angle for the only case of $Q=c$ had been estimated by using 
the HQET\cite{Ito:1996mr}.
In this paper, we estimated the mixing angle between $\varXi_Q$ and $\varXi_Q'$
for each case of $Q=c$ and $Q=b$ by using the HQET.
We confirmed the consistency of the HQET and the experiment.

\section{Mixing Angle of the $\Xi_Q-\Xi_Q'$ in Heavy Quark Effective Theory }

The HQET Lagrangian with $O\left(\Lambda_{QCD} / m_Q \right)$ corrections is
\begin{equation}
{\cal L} = {\bar h}_v(iv\cdot D)h_v+\frac{1}{2m^{}_Q}{\bar h}_v(iD)^2h_v
+Z_Q\frac{1}{2m^{}_Q}{\bar h}_v\sigma^{\mu\nu}G_{\mu\nu}h_v,
\end{equation}
where $v$ is the four velocity of the heavy quark $h$,
$Z_Q$ is a renormalization factor and $G_{\mu\nu}$
is the gluon field strength\cite{Luke:1990eg}.
Neglecting the higher order corrections $O\left(\Lambda^2_{QCD} / m^2_Q \right)$ ,
we have a heavy hadron mass formula
\begin{equation}
M\left(H_Q\right) = m^{}_Q+{\bar\varLambda}- \mql -d^{}_H\mqll, \label{eqn:mass}
\end{equation}
where ${\bar\varLambda}$ is the energy of the light degrees of freedom
in the hadron.
Two  $O\left(\Lambda_{QCD} / m_Q \right)$ correction coefficients are given by the matrix elements
\begin{equation}
\lambda_1 = \langle H_Q(v)|{\bar h}_v(iD)^2h_v|H_Q(v)\rangle
\end{equation}
and
\begin{equation}
d^{}_H\lambda_2 = \frac{1}{2}Z_Q\langle H_Q(v) |
{\bar h}_vG_{\mu\nu}\sigma^{\mu\nu}h_v |H_Q(v)\rangle ,
\end{equation}
where $d^{}_H$ is the Clebsch-Gordan factor which can be written by,
\[
d^{}_H = - 4 \langle \vec{S}_Q \cdot \vec{s}_{\ell} \rangle .
\]
$\vec{S}_Q$ and $\vec{s}_Q$ are the spins of the heavy quark and the angular momentum of
 the light degrees of freedom respectively. 
The $\langle \vec{S}_Q \cdot \vec{s}_{\ell} \rangle$ can be obtained by
\[
   2 \langle \vec{S}_Q \cdot \vec{S}_{\ell} \rangle 
   = J (J + 1) - S_Q (S_Q + 1) - S_{\ell} ( S_{\ell} + 1),
\]
where $J$, $S_Q$, $s_{\ell}$ are the magnitude of the total angular momentum and 
the spin of the heavy quark and the angular momentum of
 the light degrees of freedom respectively.
 
These baryon masses are given by\cite{Ito:1996mr}
\begin{eqnarray}
M \left(\varLambda_Q \right)&=& m^{}_Q+{\bar \varLambda}_{0^+}-\mqlt, \\
M \left(\varSigma_Q \right)&=& m^{}_Q+{\bar \varLambda}_{1^+}-\mqls -4\mqlls, \\
M \left(\varSigma_Q^* \right)&=& m^{}_Q+{\bar \varLambda}_{1^+}-\mqls +2\mqlls.\\
M \left(\varOmega_Q  \right)&=& m^{}_Q+{\bar \varLambda}''_{1^+}-\mqls -4\mqlls, \\
M \left(\varOmega_Q^* \right)&=& m^{}_Q+{\bar \varLambda}''_{1^+}-\mqls +2\mqlls .
\end{eqnarray}

The relation between heavy quark limit ${\varXi}_Q^H$(${{\varXi}'}^H_Q$)
states and physical ${\varXi}_Q$(${\varXi}'_Q$)
states can be written in terms of the mixing angle $\theta_Q$
\begin{equation}
\left(
\begin{array}{c}
|{\varXi}_Q \rangle \\
|{\varXi}'_Q \rangle
\end{array}
\right)
=
\left(
\begin{array}{cc}
\cos \theta_Q & \sin \theta_Q \\
-\sin \theta_Q & \cos \theta_Q
\end{array}
\right)
\left(
\begin{array}{c}
|{\varXi}_Q^H \rangle \\
|{{\varXi}'}^H_Q \rangle
\end{array}
\right).
\end{equation}
We may consider that the mixing angle $\theta_Q$ is a function of
$\varLambda_{\rm QCD}/m^{}_Q$ and that it becomes to zero in heavy quark limit $m^{}_Q\to \infty$.
So the masses of ${\varXi}_Q$, ${\varXi}'_Q$ and
${\varXi}_Q^*$ baryons are given by
\begin{eqnarray}
M \left(\varXi_Q \right)&=& m^{}_Q +\cc\;{\bar \varLambda}'_{0^+} +\sss\;{\bar \varLambda}'_{1^+} \nonumber \\
  & & -\cc\;\mqlt-\sss \left(\mqls -4\mqlls \right), \\
M \left(\varXi'_Q \right)&=& m^{}_Q +\sss\;{\bar \varLambda}'_{0^+} +\cc\;{\bar \varLambda}'_{1^+} \nonumber \\
 & & -\sss\;\mqlt -\cc \left( \mqls -4\mqlls \right), \\
M \left(\varXi_Q^* \right)&=& m^{}_Q + {\bar \varLambda}'_{1^+} -\mqls +2\mqlls.
\end{eqnarray}
The mixing angle $\theta_Q$ can be expressed in terms of the baryon masses,
\begin{equation}
\sss = \frac{M \left(\varXi'_Q \right) -M \left(\varXi_Q^* \right) + M \left(\varSigma_Q^* \right) -M \left(\varSigma_Q \right)}
{M \left(\varXi_Q \right)+M \left(\varXi'_Q \right)-2M \left(\varXi_Q^* \right) 
+ 2M \left(\varSigma_Q^* \right)-2M \left(\varSigma_Q \right)
}. \label{eqn:angle}
\end{equation}

To estimate these mixing angles, $\theta_c$ and $\theta_b$, we use experimental values of the baryon masses in Table \ref{baryonmass}.
\begin{table}[htbp]
\begin{center}
\begin{tabular}{|c|c|c|}
\hline
 $M\left( H_Q \right)$ [MeV] & $c$ & $b$\\
\hline
 $\varSigma_Q$ &  $2453.54 \pm  0.15 $ & $5813.10 \pm 0.18 $  \\
\hline
 $\varSigma^*_Q$&  $2518.45 \pm 0.15 $ &   $5832.53 \pm  0.20$\\
\hline
$\varXi_Q$ & $2469.42 \pm 0.15 $ & $5794.45 \pm 0.51 $  \\
\hline
 $\varXi'_Q$ & $2578.80 \pm 0.35  $ &  $ 5935.02 \pm 0.05 $ \\
\hline
 $\varXi^*_Q$&  $ 2645.98 \pm 0.17  $  &   $ 5955.33 \pm  0.13$ \\
\hline
\end{tabular}
\end{center}
\caption{The experimental values of the baryon masses including a heavy quark\cite{PDG2020}. These values are averaged about the electric charge, 
if there are multiple status which are different about the electric charge.}
\label{baryonmass}
\end{table}

Our results of the mixing angle between $\varXi_Q$ and $ \varXi'_Q$ baryons are:
\[
\left|\theta_c\right| = 8.12^{\circ} \pm 0.80 ^{\circ}
\]
for the  $\varXi_c$ - $\varXi'_c$ mixing and
\[
\left|\theta_b\right| = 4.51^{\circ} \pm 0.79 ^{\circ}
\]
for the  $\varXi_b$ - $\varXi'_b$ mixing.
By using the mass relations, we can't determine the sign of the mixing angle.
In these results, there are errors which originates in the experimental ones. 
In addition, these results contain the uncertainties due to neglecting
the next to order corrections  $O\left(\Lambda^2_{QCD} / m^2_Q \right)$.

\section{Summary and Discussion}

It was found that the mixing angle between 
$\varXi_Q$ and $varXi_Q'$ is given by 
$\left|\theta_b \right|= 4.51^{\circ} \pm 0.79 ^{\circ}$ for $Q = b$ case and
$\left|\theta_c\right| = 8.12^{\circ} \pm 0.80 ^{\circ}$ for $Q = c$ case.

In Table \ref{mixingangles}, we summarize the values predicted by other methods, QCD sum rules\cite{Aliev:2010ra} and 
quark model\cite{Franklin:1981rc,Franklin:1996ve}. 
It is seen, within errors, our estimation for $\theta_c$ is in agreement with the predictions by
QCD sum rules,
whereas, the others are in disagreement.

\begin{table}[htbp]
\begin{center}
\begin{tabular}{|c|c|c|c|}
\hline
 $\left|\theta_Q\right|$ & This Work  & QCD sum rules  & quark model\\
\hline
 $\left|\theta_c\right|$& $8.12^{\circ} \pm 0.80 ^{\circ}$ &  $5.5^{\circ} \pm 1.8 ^{\circ}$ & $3.8 ^{\circ}$\\
\hline
 $\left|\theta_b\right|$& $4.51^{\circ} \pm 0.79 ^{\circ}$ & $ 6.4^{\circ} \pm 1.8 ^{\circ}$  & $ 1.0 ^{\circ}$\\
\hline
\end{tabular}
\end{center}
\caption{Our estimation of the mixing angles along with the predictions of QCD sum rules\cite{Aliev:2010ra} and quark model\cite{Franklin:1981rc,Franklin:1996ve}.}
\label{mixingangles}
\end{table}

In heavy quark limit $m_Q \rightarrow \infty$, the mixing angle
between $\varXi_Q$ and $ \varXi'_Q$ baryons should be zero. 
Therefore, it is expected that the mixing angle should be small,
so that  quark mass is heavy.  
Our results agreed with this expectation.

\end{document}